\documentstyle[12pt,epsfig]{article}
\newcommand{\be}{\begin{equation}}
\newcommand{\ee}{\end{equation}}
\newcommand{\bd}{\begin{description}}
\newcommand{\ed}{\end{description}}
\newcommand{\bmat}{\begin{displaymath}}
\newcommand{\emat}{\end{displaymath}}
\newcommand{\bit}{\begin{itemize}}
\newcommand{\eit}{\end{itemize}}
\newcommand{\ben}{\begin{enumerate}}
\newcommand{\een}{\end{enumerate}}
\begin{document}
\begin{flushright}
PRA--HEP/94--11
\end{flushright}
\vspace*{1cm}

\begin{center}
{\bf ON THE CHARGE DEPENCENCE OF FACTORIAL MOMENTS IN $e^+e^-$
PROCESSES}
\footnote{Talk presented at the XXIV International Symposium
on Multiparticle Dynamics, Vietri sul Mare, September 12--19 1994.} \\
\vspace*{1.0cm}
Ji\v{r}\'{\i} Rame\v{s} \\
\vspace*{0.3cm}
{\em Institute of Physics, Academy of Sciences of the Czech Republic  \\
Na Slovance 2, Prague 8, 18040 Czech Republic} \\
\vspace*{0.8cm}
 ABSTRACT    \\
\vspace*{0.5cm}
\parbox{5in}{\small The behaviour of like-charge and unlike-charge
second factorial
moments in $e^+e^-$ reactions is discussed. It is argued that Monte Carlo
calculation with the help of generators JETSET 7.4 and HERWIG 5.8 points
to the conclusion that the only nontrivial cause of intermittency are
Bose-Einstein correlations of identical particles. }
\end{center}

\vspace*{1cm}
\noindent

Since the original proposal of Bialas and Peschanski \cite{BiaP},
factorial moments (as well as more general quantities of the same
character \cite{Lipa})
have become a widely accepted tool in studying multiparticle
final states in various processes. Let us define the i-th factorial
moment as
\be
F_{i}=\frac{1}{N_{events}}\sum_{events}
\frac{\sum_{k=1}^{n_{bins}} \left\{n_{k}(n_{k}-1)
\cdots (n_{k}-i+1)\right\}/n_{bins}}
{(\langle n\rangle /n_{bins})^{i}}
\label{genfacmom}
\ee
where $\langle n \rangle$ is the average number of particles in the full
phase space region accepted, $b_{bins}$ denotes the number of bins in
this region, which is given by $(2^b)^d,\;b=0,1,2...$ ($d$ is the
dimension of the phase space region considered) and $n_k$ is the
multiplicity in $k$-th bin. In what follows I will consider factorial
moments in two and three phase-space dimensions, in the conventional
variables $(y,\varphi)$ and $(y,\varphi,\tilde{p_t})$ ($y$ denotes rapidity,
taken here from the interval $(-3.2,3.2)$, $\varphi$ is the azimuthal
angle and $\tilde{p_t}$ is the ``flattened'' momentum transverse defined as
\cite{ptilda}
\be
\tilde{p_t}=\frac{\int_{0}^{p_t}P(p)dp}{\int_{0}^{p_t^{max}}P(p)dp}
\label{ptilda}
\ee
where $P(p_t)$ is the probability distribution of $p_t$ in the interval
$(0,p_t^{max})$ and $p_t^{max}$ was chosen as $2$ GeV/c. Rapidity and $p_t$
are expressed with respect to the thrust axis. The moments are
calculated from the charged particles in final states only.

The behaviour of factorial moments plotted as a function of bin size
(i.e. $b$ in our notation)
provides information about the character of multiplicity
fluctuations among different bins. Rising of $F_i$ with rising $b$
(decreasing bin size) generally signalizes deviation from purely
Poissonian distribution of fluctuations. The linear growth
of $\log F_i$ with $b$ was called intermittency by
the authors of \cite{BiaP}. Nowadays, the term is often used for any
type of growth of $F_i$ observed.

The purpose of this talk is to discuss the effect of intermittency in
$e^+e^-$ physics in the context of Monte Carlo (MC) generators. MC
event generators provide a computational device in a situation where
one based purely on the first principles is lacking. They use
a combination of theoretically based calculations on the
level of partons and purely phenomenological (though with sound qualitative
theoretical motivation) description of the hadronization phase.
Obviously, MC generators are not physics, but for the present
discussion they have one important advantage ---
they allow one to switch on and off different particular
processes, both on the level of partons and on the level of
hadronization and hadronic decays, and thus to pinpoint in greater detail the
causes of different effects --- something which is not possible in the
real world.

\begin{figure}
\begin{center}
\epsfig{file=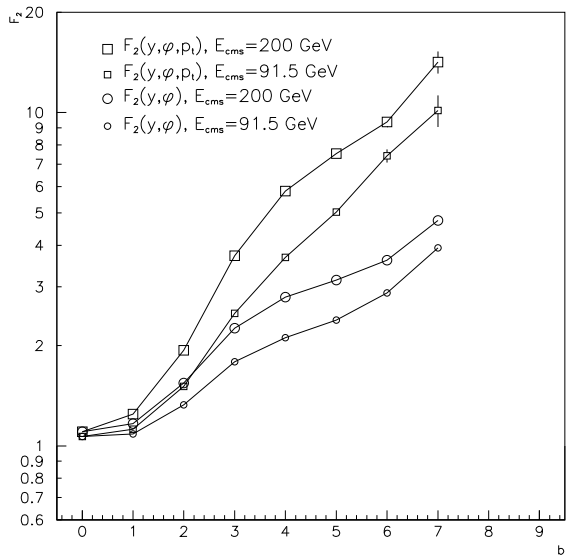,height=7cm}
\end{center}
\caption{Factorial moments $F_2$, for two and three phase space variables
 (see text), calculated by JETSET 7.4 for C.M.S. energy $91.5$ GeV and
 $200$ GeV}
\end{figure}

In what follows I will discuss the results obtained with the help of
two of the most widely used $e^+e^-$ MC event generators --- JETSET 7.4
\cite{Jetset} and HERWIG 5.8 \cite{Herwig}. The aim is to find
out what can be said about the behaviour of factorial
moments on the basis of MC generation of events and, also, how the
results of both abovementioned generators compare.

The event generators for $e^+e^-$ physics are generally considered to be
in a very good shape.
Unlike in other types of reactions, the agreement with the real data
had been reported not only for various one--particle distributions, but,
in some cases, also for multiparticle quantities like
factorial moments. At first sight, this seems not
to be surprising, as in $e^+e^-$
processes
everything is much ``cleaner'' than in other cases (no need of introducing
phenomenological parton distributions in the initial states, no
``nonperturbative'' low $p_t$ hadronic interactions etc.). On the other
hand, if, as is now increasingly believed, the major part of
intermittent behaviour is solely due to Bose-Einstein (BE) correlations
\cite{HBT} of identical particles (for a rewiev see
e.g. \cite{BialasVie}), this observation deserves some other discussion
---  as there are no true BE correlations included in MC generators
 (more on this see below).

Both of the generators mentioned above use a QCD motivated
description of the partonic stage of the process, so that it
is possible to address, at least in principle, the question whether
intermittency is due to some sort of partonic branching (as, for
example, descriptions based on local parton-hadron duality (LPD)
\cite{LPD} claim), or whether it is the result of some other mechanism.
The question had been partially answered in \cite{My}, where it is
stated that there is no LPD present either in JETSET or in HERWIG ---
although the generators are  considerably different on the level of partons,
in terms of final hadrons they yield almost identical results not only
for various one-particle distributions, but also rising behaviour for
factorial moments. In this talk, I will discuss this problem further.

\begin{figure}
\begin{center}
\epsfig{file=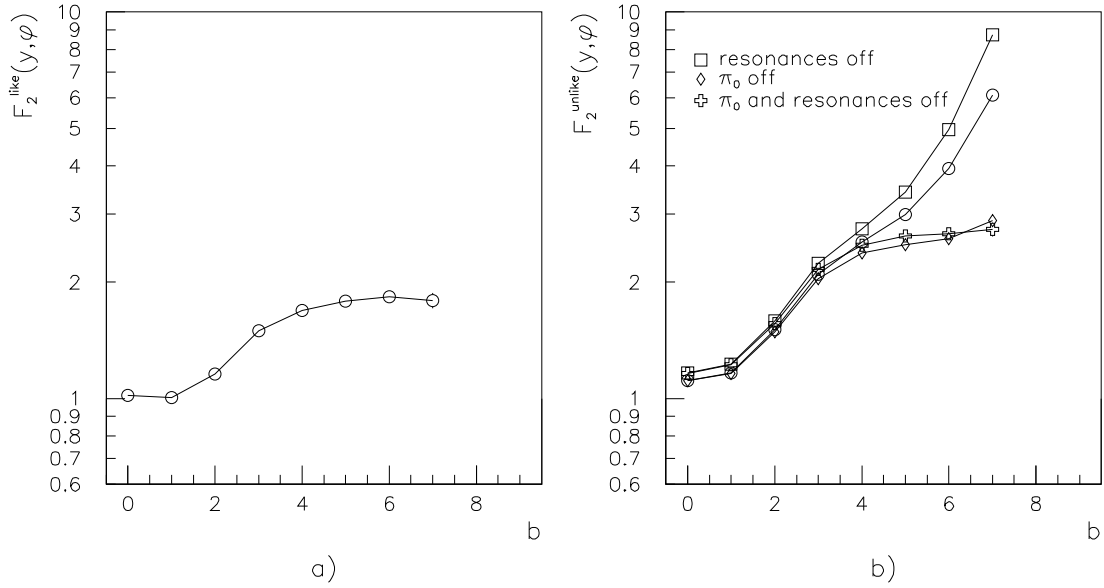,height=7cm}
\end{center}
\caption{Factorial moments $F_2(y,\varphi)$
 for like-charge (a) and unlike-charge (b)
 combinations, calculated by JETSET 7.4 for C.M.S. energy $91.5$ GeV.
 Figure (b) illustrates the influence of resonance decays and $\pi_0$
 decays.}
\end{figure}

Fig.1 shows the second factorial moment $F_2$, both for two-dimensional
and for three-dimensional case, calculated with the help of JETSET 7.4,
as a function of bin size characterized by $b$. In both cases one
observes clear rising behaviour. The question is what is the cause of
this behaviour: One can be sure (in contrast to real data) that there
are no BE effects included in the MC calculation, while direct effect of
partonic branching is strongly disfavored \cite{My}.
It is clear that more information can be
obtained if one looks at the moments for like-charge and
unlike-charge combinations of particles separately ---
in the like-charge quantities the BE correlations
may play their role (though, I repeat, not in the MC
calculations presented here), while in an unlike-charge
factorial moment the influence
of BE correlation is principially excluded. The second
factorial moments for like charge pairs are defined in an obvious way --- e.g.,
for plus-plus combination as
\be
F_{2}^{++}=\frac{1}{N_{events}}\sum_{events}
\frac{\sum_{k=1}^{n_{bins}} \left\{n_{k}^{+}(n_{k}^{+}-1)
\right\}/n_{bins}}
{(\langle n^{+}\rangle /n_{bins})^{2}}
\label{Fstejny}
\ee
and analogously for minus-minus. For unlike charges I use the definition
\be
F_{2}^{+-}=\frac{1}{N_{events}}\sum_{events}
\frac{\sum_{k=1}^{n_{bins}} \left\{n_{k}^{+}n_{k}^{-}
\right\}/n_{bins}}
{\langle n^{+}\rangle \langle n^{-}\rangle /(n_{bins})^{2}}
\label{Fruzny}
\ee
Fig.2 shows the behaviour of  two-dimensional
factorial moments at the centre-of-mass energy $91.5$ GeV,
calculated with the help of JETSET generator.
One can see that the like-charge moment shows a clear plateau
(if BE correlations are the real cause of intermittency, one should
expect the rise of this quantity for the real data --- a conclusion
\begin{figure}
\begin{center}
\epsfig{file=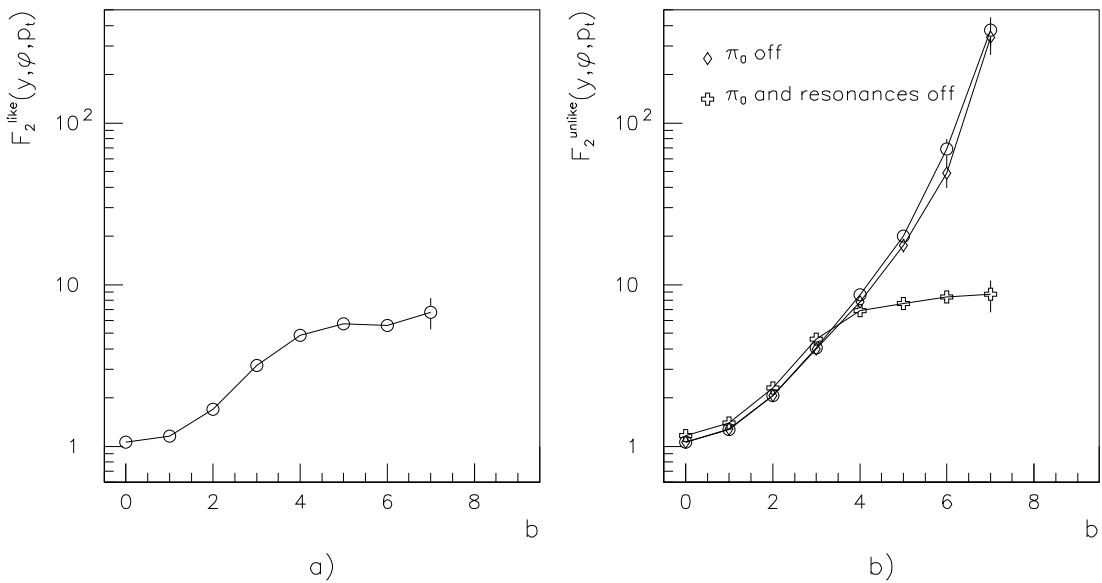,height=7cm}
\end{center}
\caption{Factorial moments $F_2(y,\varphi,\tilde{p_t})$ for like-charge (a)
 and unlike-charge (b) combinations, calculated by JETSET 7.4 at
 C.M.S. energy $200$ GeV.}
\end{figure}
supported e.g. by \cite{Amiran} (which, however, analyses not factorial
moments but two-particle correlation functions), while all the observed rising
behaviour is in the unlike-charge function.

An obvious candidate to cause this type of behaviour is the
decay of resonances in the unlike-charge channel. However, the
line on Fig.2b, corresponding to the calculation with several
major resonances ($\rho_0,\eta,\eta',\eta_c,\omega,B,J/\psi$)
switched off in the generator, shows that this is not
the case. On the other hand, there is a clear plateau on the line
corresponding to $\pi_0$ decays switched off. The only important
unlike-charge decay products of $\pi_0$ decay are $e^+e^-$ Dalitz pairs.
The rising of unlike-charge two-dimensional second order factorial
moments (as well as that part of all-charge moments rising which is not
caused by BE correlations) is thus due to Dalitz pairs.  The line
in Fig.2b corresponding to both $\pi_0$ and resonance decays switched
off shows that for two-dimensional moments, resonance decays do not
play any important role in the phenomenon of intermittency.

The same quantities, but this time for the three-dimensional case in the
variables $y$, $\varphi$ and $\tilde{p_t}$, are depicted in Figs.3a,
3b. As Fig.3a shows, for like-charge FM the situation is much the same
as for two dimensions. For the unlike-charge combinations, there is,
however, one difference --- to ``stop'' the rise of the quantity, one has to
switch off both $\pi_0$ decays and resonances. This is another
demonstration of the well known fact that the most detailed information
is provided by three-dimensional factorial moments, while in lower
dimensions things may ``project out''.

\begin{figure}
\begin{center}
\epsfig{file=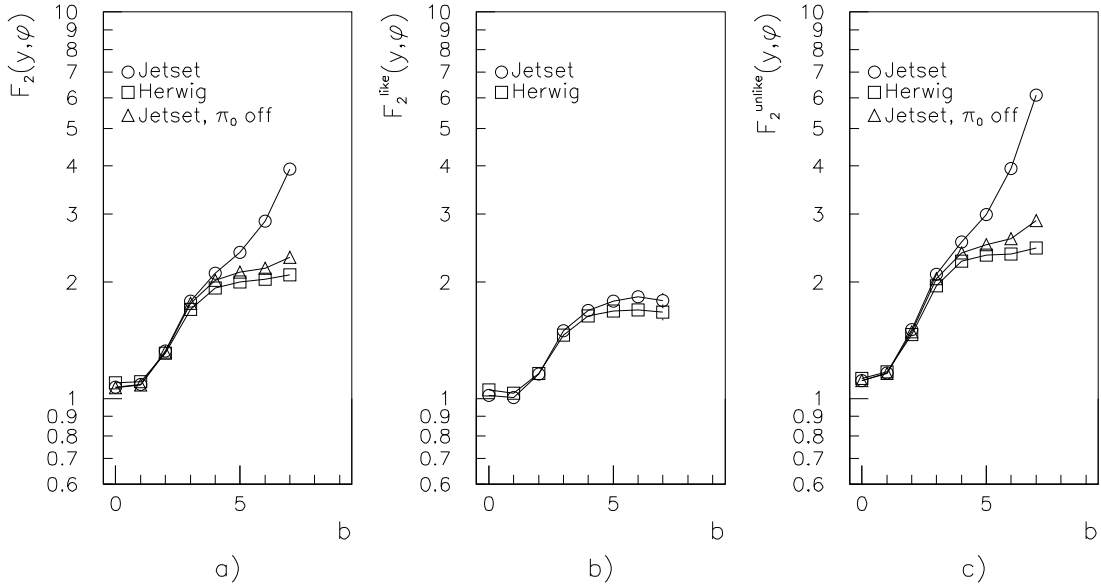,height=7cm}
\end{center}
\caption{Comparison of two-dimensional factorial moments $F_2$
 calculated by JETSET 7.4 and HERWIG 5.8 for $E_{CMS}=91.5$ GeV}
\end{figure}

To summarize this part, from the analysis based on the calculations
with the generator JETSET 7.4 it follows that the only cause of
the intermittent behaviour of factorial moments, with the possible
(and probable) exception of BE effects, are various decays of
known hadrons. How do the predictions of HERWIG 5.8 look like in
this respect?

Fig. 4a shows a comparison, for the two-dimensional moment, between
a calculation done by JETSET and HERWIG, respectively.
There is a clear difference --- the curve corresponding to HERWIG
shows a plateau. In view of the previous experience,
however, one can suspect that the situation looks very much
like in JETSET with $\pi_0$ decays switched off. Indeed, one finds out
that the decay mode of $\pi_0$ into $e^+e^-$ pair
is not included in the hadronic decays of HERWIG
5.8. If this situation is ``corrected'' by switching of $\pi_0$ decay
in JETSET, a good agreement between
the predictions of the two generators is obtained
(fig. 4a, line of triangles).
Figs. 4b and 4c show the comparison of JETSET and HERWIG predictions
for like-charge and unlike charge moments separately. They confirm
the conlusion that the only (but, as we have seen, importatnt
for the behaviour of factorial moments) disagreement is in the
non-inclusion of Dalitz decays of $\pi_0$ mesons in HERWIG. Otherwise,
both generators give very similar results, much the same way as for the
one-particle distributions.

\begin{figure}
\begin{center}
\epsfig{file=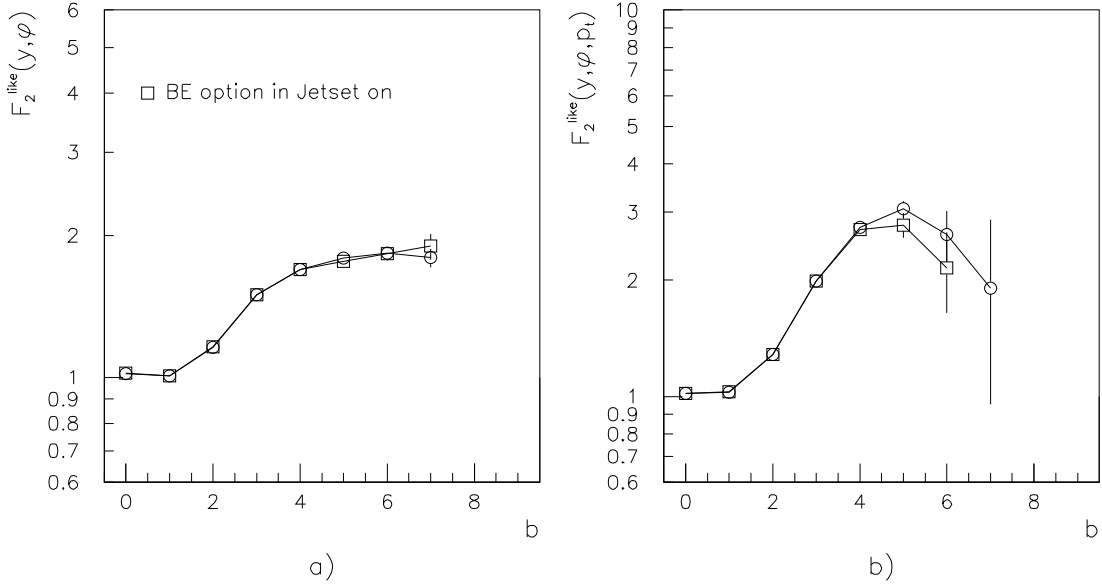,height=7cm}
\end{center}
\caption{Factorial moments $F_2$ for unlike-charge combinations,
 calculated by JETSET 7.4 without and with the Bose-Einstein option.
 $E_{CMS}=91.5$ GeV}
\end{figure}

The same conclusion follows from the comparison of three-dimensional
factorial moments (not shown here).

In the JETSET generator, there is an option to include the effect of
 BE correlations.
Of course, no generator can incorporate true BE correlations, as
generators do not describe coherent effects in hadronization.
In JETSET, the BE option consists in
modification of two-particle distribution of identical particles
in the $Q^2$ variable (where $Q^2 = -(p_i - p_j)^2$) according to
certain weight. It is nevertheless interesting to find out how
the factorial moments from JETSET look like with BE correlations
option switched on. This is shown on Fig.5: The JETSET option
of including BE correlations has no apparent influence on the character
of behaviour of like-charge factorial moment.
This is seemingly in
disagreement with what is stated in some analyses.
The whole point, however, is that the analyses which
report increased agreement between MC generation and measured quantities
after inclusion of BE option in JETSET (e.g. \cite{Amiran})
are precisely those which use
variables like $Q^2$ or $m^2$ (the invariant mass
of the pairs). Throughout this discussion, however, I used the "old
fashioned" way of expressing factorial moments in separate phase-space
variables. It is not surprising that the method of inclusion BE
correlations which is tailored to the $Q^2$ variable works reasonably
well with functions of this same variable but fails in other cases.
As mentioned above, truly general description of BE correlations within
the generators is, up to this moment, lacking.

In summary, the present analysis leads to following conclusions:
\bit
\item The calculation of two- and three-dimensional second factorial
moments based on the MC event generators JETSET 7.4 and HERWIG 5.8
supports the view that all the intermittency (i.e. rising behaviour of
factorial moments) in $e^+e^-$ reactions is due to either BE
correlations of identical hadrons or known effects as decays of
particles. Especially important is the role of decays of $\pi_0$ mesons
into $e^+e^-$ Dalitz pairs. No new mechanism is present.
\item The
generators used for the calculation yield very similar results (apart
from the trivial technical detail that a decay mode important for the
effect studied, the $\pi_0$ decay into Dalitz pairs, is not included in
HERWIG). Apart from the importance of the fact itself, this leads to an
immediate conclusion concerning local parton hadron duality: One can
hardly imagine that something like local parton hadron duality is
present in the generators used, as both of them differ quite
considerably on the parton level \cite{My}. In a sense, the
hadronisation in the generators considered has just the opposite property,
being highly nonlocal.
\item The effect of BE option in JETSET on the calculated
quantities is highly dependent on the type of quantities and,
especially, variables used. In the case discussed (factorial moments as
functions of separate non-invariant phase-space variables) the inclusion
of BE option shows little influence on the character of behaviour
of factorial moments.
This fact is hardly surprising, in view of how BE effects in JETSET are
treated \cite{Jetset}.
\eit


\end{document}